\documentclass[12pt]{iopart}

\usepackage{iopams}
\usepackage{epsfig}
\begin{document}
\def\be {\begin{equation}}
\def\ee {\end{equation}}
\def\nn {\nonumber}
\def\bea {\begin{eqnarray}}
\def\eea {\end{eqnarray}}
\def\k {|{\vec k}|}
\def\q {|{\vec q}|}
\def\ks {{|{\vec k}|}^\ast}
\def\qs {|{{\vec q}|}^\ast}
\def\mqs {M_q^\ast}
\def\eqs {E_q^\ast}
\def\eq {E_q}
\def\mks {M_k^\ast}
\def\ek {E_k}
\def\eks {E_k^\ast}
\def\mp {m_\pi}
\newcommand{\bef}{\begin{figure}}
\newcommand{\eef}{\end{figure}}
\newcommand{\ra}{\rightarrow}
\newcommand{\gm}{\gamma^\mu}
\newcommand{\gf}{\gamma^5}
\newcommand{\N}{\bar{N}}
\newcommand{\del}{\partial}
\def\bt{{\bf{\tau}}}
\def\ba{{\bf{a_1}}}
\def\br{{\bf{\rho}}}
\def\bp{{\bf{\pi}}}
\def\cl{{\cal{L}}}
\title[]{Thermal photons to dileptons ratio at LHC}

\author{Jajati K. Nayak, Jan-e Alam, Sourav Sarkar and Bikash Sinha}

\address{Variable Energy Cyclotron Centre, 1/AF Bidhan Nagar,
Kolkata 700 064, INDIA} 

Photons and dileptons are  considered to be efficient
probes of quark gluon plasma (QGP) expected to be created  
in heavy ion collisions at ultra-relativistic
energies.  However, the theoretical calculations of
the transverse momentum ($p_T$) spectra of photons 
($d^2N_\gamma/d^2p_Tdy_{y=0}$)
and dileptons ($d^2N_{\gamma^\ast}/d^2p_Tdy_{y=0}$)
depend on several parameters which are model dependent 
(see~\cite{nayak,alam} and 
references therein). 
In the present work it is shown that the 
model dependences  involved in 
individual photon and dilepton 
spectra are canceled out in the ratio, $R_{em}$ defined as:
$R_{em} = (d^2N_\gamma/d^2p_Tdy)_{y=0}/(d^2N_{\gamma^\ast}/d^2p_Tdy)_{y=0}$.

The invariant yield of thermal photons can be written as 
${d^2N_\gamma}/{d^2p_{T}dy}=\sum_{i=Q,M,H}{\int_{i}{\left({d^2R_\gamma}/
{d^2p_{T}dy}\right)_i d^4x}}$,  
where $Q, M$ and $H$ represent QGP, mixed (coexisting
phase of QGP and hadrons)
and hadronic phases respectively.
$(d^2R/d^2p_{T}dy)_i$ is the static rate of photon
production from the phase $i$, which is convoluted over
the expansion dynamics
through the integration over $d^4x$.
The thermal photon rate from QGP up to 
$O(\alpha \alpha_{s})$ have been considered.  For photons from 
hadronic matter an exhaustive set of reactions
(including those involving strange 
mesons) and radiative decays of higher resonance states 
have been considered in which form factor effects have been
included.

Similar to photons, the $p_T$ distribution 
of thermal dileptons is given by,  
${d^2N_{\gamma^\ast}}/{d^2p_{T}dy}=\sum_{i=Q,M,H}{\int_{i}
{\left({d^2R_{\gamma^\ast}}/{d^2p_{T}dydM^2}\right)_i dM^2d^4x.}}$
The limits for the integration over $M$ are fixed  
from experimental measurements. Here  we consider
$2m_{\pi} < M $ $ <1.05$ GeV. 
Thermal dilepton rate from QGP 
up to $O(\alpha^2 \alpha_{s})$ has been considered. 
For the hadronic phase we include the dileptons 
from the decays of light vector mesons
~\cite{nayak}. The space time evolution 
of the system has been studied using 
(2+1) dimensional  relativistic hydrodynamics 
with longitudinal boost invariance and cylindrical 
symmetry. 
The calculations have been performed for the initial conditions
mentioned in table 1 (see also~\cite{nayak}). 
The values of parameters shown in table 1 reproduce
the various experimental data from SPS and RHIC. 
For LHC we have chosen two values of $T_i$
corresponding to two values of $dN/dy$.
We use the Bag   model EOS
for the QGP phase. For  EOS of the hadronic matter
all resonances with mass $\leq 2.5$ GeV have been considered

\begin{figure}
\begin{flushleft}
\includegraphics[width=8.0cm]{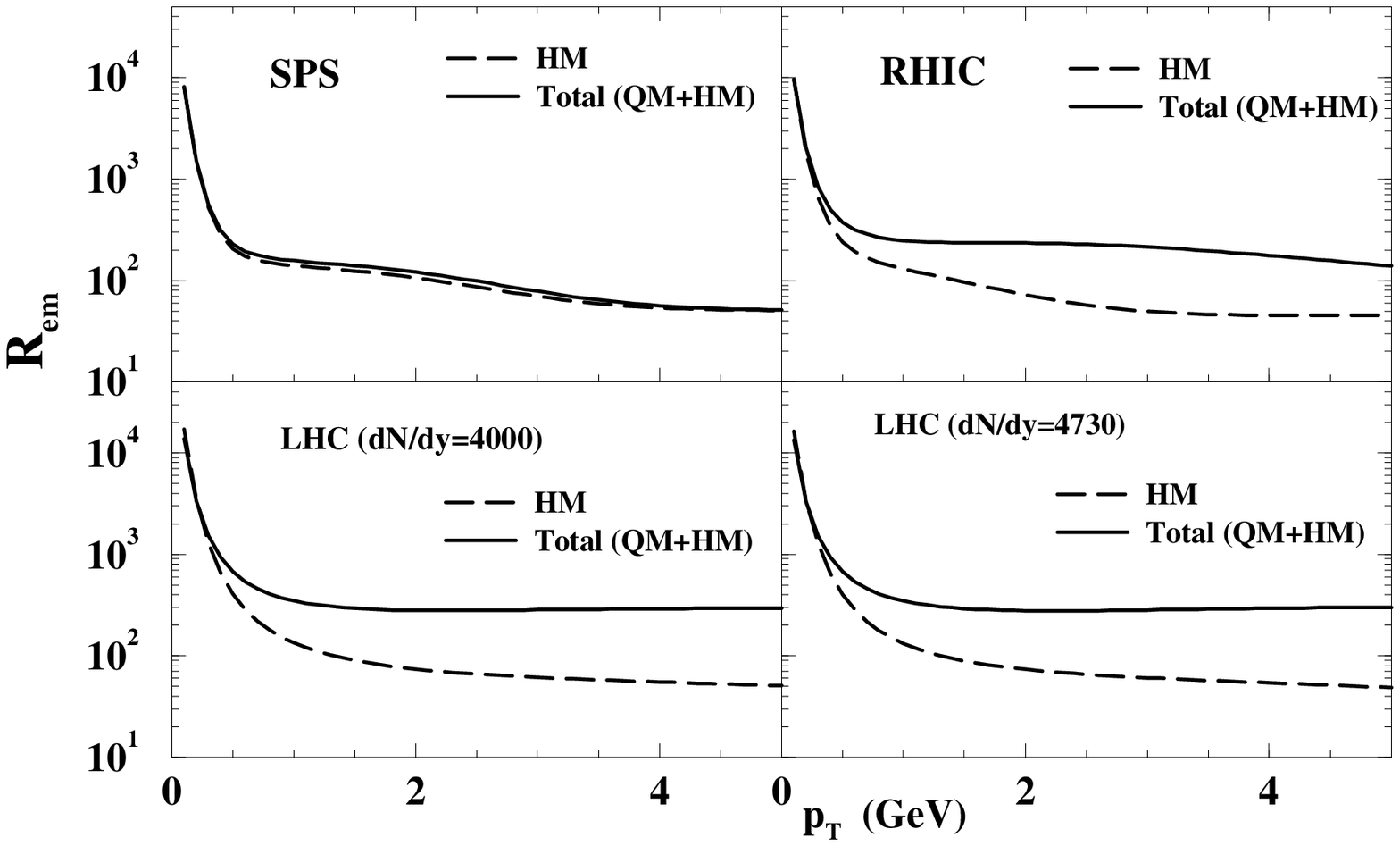}
\end{flushleft}
\begin{flushright}
\vspace{-5.5cm}
\includegraphics[width=6.3cm]{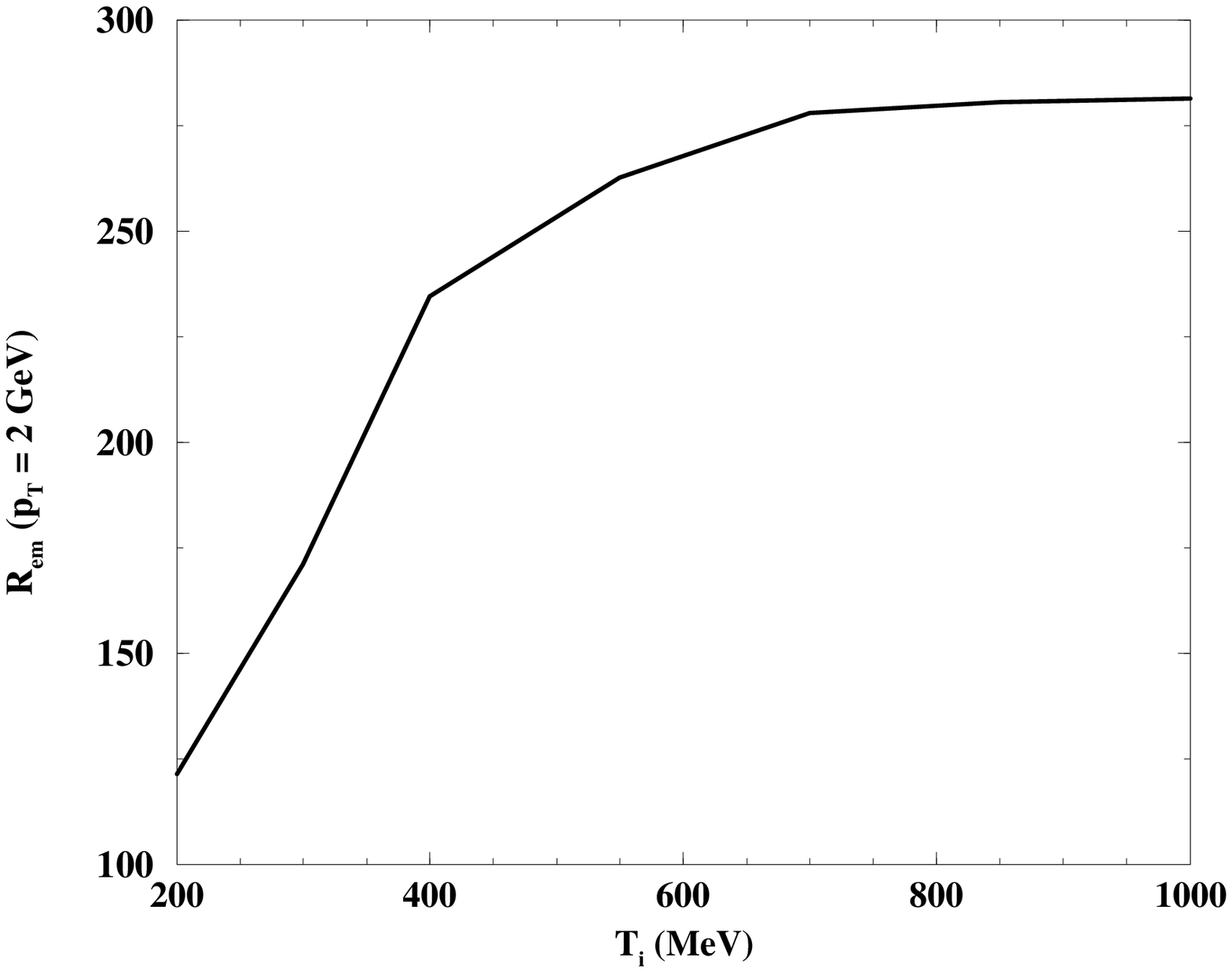}
\end{flushright}
\hspace{-12.0cm}{\caption{Left panel: Variation of $R_{em}$ with $p_T$,
right panel: variation of $R_{em} (p_T=2$GeV) with $T_i$.}}
\label{fig1}
\end{figure}
\begin{table}
\caption{The values of various parameters - thermalization
time ($\tau_i$), initial temperature ($T_i$), freeze-out temperature
($T_f$) and hadronic multiplicity $dN/dy$  - used
in the present calculations.}
\begin{tabular}{lcccr}
Accelerator&$\frac{dN}{dy}$&$\tau_i(fm)$&$T_i(GeV)$ &$T_f (MeV)$\\
SPS&700&1&0.2&120\\
RHIC&1100&0.2&0.4&120\\
LHC&4000&0.08&0.85&120\\
LHC&4730&0.08&0.905&120\\
\end{tabular}
\end{table}
The variation of $R_{em}$ with $p_T$ for different
initial conditions are  
depicted in Fig.~1 (left panel). 
At SPS, the contributions from hadronic matter (HM) coincides with
the total and hence it becomes difficult to make any
conclusion about the formation of QGP. However,
for RHIC and LHC the  contributions from HM are
less than the total indicating large contributions from
quark matter. The quantity, $R_{em}$, reaches a plateau 
beyond $p_T=1$ GeV
for all the three cases {\it i.e.} for SPS, RHIC and LHC.
However, it is very important to note that the values
of $R_{em}$ at the plateau region are different,
{\it e.g.} $R_{em}^{LHC}\,>\,R_{em}^{RHIC}\,>\, R_{em}^{SPS}$.
Now for all the three cases, SPS, RHIC and LHC, except
$T_i$  all other quantities {\it e.g.} $T_c$, $v_0$, $T_f$ and EOS
are same, indicating that  the difference in the value of $R_{em}$ 
in the plateau
region originates only due to different values of $T_i$ for the 
three cases (Fig.1, right panel).
This, hence can be used as a measure of $T_i$.

We have observed that although the individual $p_T$ distribution 
of photons and lepton pairs are sensitive to different
EOS (lattice QCD, for example) the ratio $R_{em}$ is not.
It is also noticed that $R_{em}$ in the plateau 
region is not sensitive to the medium effects on hadrons, 
radial flow, $T_c$, $T_f$ and  other parameters.

It is interesting to note that the nature of variation of the
 quantity, $R_{em}^{pQCD}$, which is the corresponding ratio of
photons and lepton pairs from  hard processes only is
quite different from  $R_{em}^{thermal}$ for $p_T$ up to $\sim 3 $ GeV
indicating that the observed saturation is a thermal effect.

\end{document}